\documentclass[jap,preprint,preprintnumbers,amsmath,amssymb]{revtex4}

\usepackage[T1]{fontenc}
\usepackage[american]{babel}
\usepackage{graphicx}			
\usepackage{dcolumn}			
\usepackage{bm}					
\usepackage{SIunits}
\usepackage{natbib}

\newcommand{\etal}{\textit{et al.}}
\newcommand{\ie}{\textit{i.e.}}

\newcommand{\LNO}{LaNiO$_3$}
\newcommand{\LNOo}{LaNiO$_{3-\delta}$}

\begin{document}

\title{Redox reaction enhanced Schottky contact at a \LNO{}(001)/Al interface}

\author{J. Scola}
\email{joseph.scola@uvsq.fr}
\affiliation{Groupe d'Etude de la Matière Condensée (GEMaC), Université de Versailles Saint-Quentin-en-Yvelines, CNRS-UMR8635, Université Paris-Saclay, 78035 Versailles, France}
\author{B. Berini, Y. Dumont}
\affiliation{Groupe d'Etude de la Matière Condensée (GEMaC), Université de Versailles Saint-Quentin-en-Yvelines, CNRS-UMR8635, Université Paris-Saclay, 78035 Versailles, France}
\author{P. Nukala, B. Dkhil}
\affiliation{Laboratoire Structures, Propriétés et Modélisation des Solides, CentraleSupélec, CNRS-UMR8580, Université Paris-Saclay, 91190 Gif-sur-Yvette, France}


\begin{abstract}
Emergent phenomena at interfaces between oxides and metals can appear due to charge transfer and mass transport that modify the bulk properties.
By coating the metallic oxide LaNiO$_3$ by aluminium, we fabricated a junction exhibiting a diode-like behaviour.
At the equilibrium, the interface is insulating.
The metallic behaviour can be recovered by applying a voltage drop across the junction in one polarity only.
The electrical properties in direct and reverse bias are investigated.
The observed electro-resistive effect rises up to $10^5$ \% and can be interpreted in terms of (i) a spontaneous redox reaction occurring at the interface and (ii) its reversal induced by charge injection in direct bias.
\end{abstract}

\maketitle

\section{Introduction}
Oxide surfaces and interfaces attract a growing attention due to the emergence of several phenomena which are absent in the bulk.
Progress in synthesis of thin films enabled to get rid of extrinsic factor like surface roughness or grain boundaries and the engineering of heterostructure of ultra-thin films has become a new way to manipulate the electronic structure at the surfaces and interfaces.

Nickelates offer an ideal background for the emergence of surface and interface phenomena :
(i) their ground state is close to a metal-to-insulator transition (MIT) that occur between 100 and 600 K depending of the rare earth ionic radius (except \LNO{} that remains metallic at the lowest temperature) \cite{garciamunoz1993, zhou2003};
(ii) nickel sites are magnetic and order at low temperature \cite{guo2018};
(iii) their electronic structures are strongly coupled with structural distortions due to their electron correlations;
(iv) the valence band is degenerated and has a hybridized Ni and O character that makes them charge transfer insulators in the insulating side of the MIT :
hence, not only does oxygen vacancy distort the ionic lattice but it directly affect the filling of the valence band as well \cite{zaanen1985}.

Proper epitaxial strain have proved to lift the orbital degeneracy \cite{chakhalian2011, nowadnick2015, he2015} resulting in orbital polarization of the electronic structure \cite{peil2014}.
The band structure can also be tailored by the breaking of the translational invariance at a surface or an interface causes a charge redistribution thus creating an electronic structure different from that of the bulk \cite{disa2015}.
Surface termination layer can be chosen to adjust the surface electronic state via 
surface electrostatic properties \cite{kumah2014}.
Emergent surface magnetic order at the interface between non magnetic compounds have been reported too \cite{gibert2012}.

In the case of an metal-oxide interface, mass transport can occur and modify the charge transfer. 
If the mass transport takes place over more than one monolayer, the situation can be described in terms of a chemical interaction (redox reaction, alloy formation, encapsulation or interdiffusion) yielding new compounds that intervene in the charge redistribution \cite{fu2007}.
This principle has been recently applied in a universal method to fabricate 2D electron systems with insulating oxides \cite{roedel2016}.
Used on metallic \LNO{} this method leads to an electro-resistive effect reported in Ref. \cite{tian2017}.

Here, we describe the reduction of an oxide by an aluminium overlayer deposited on a the metallic oxide \LNO{}.
The latter compound hosts a metal-to-insulator transition as oxygen vacancies are introduced in the material \cite{berini2007prb, berini2008}.
The electric properties of the individual compounds of the Al/\LNO{} interface are  dramatically modified by the solid phase redox reaction that take place.
The reaction can be explained in terms of transport of oxygen ions through the interface.
We show that the reaction can be reversed by an external electric field and the junction resistance varies by a factor up to 1000.

\section{Experimental procedure}

The \LNO{} films were epitaxially grown on (001) SrTiO$_3$ substrates by pulsed laser deposition  and subsequently annealed during 45 min under oxygen.
The growth procedure and the quality of the film are reported in Ref. \cite{berini2007, berini2008}
The oxygen stoichiometry of the LaNiO$_{3-\delta}$ films could be estimated by a direct procedure described in \cite{scola2017} and the average $\delta$ is found to be less than 0.2 here.
A 80 nm thick Al overlayer was then deposited by using a rf-sputtering on three \LNO{} films having three different thickness (6, 10 and 16 nm).

The electrical properties were probed in a standard 2-wire configuration.

\section{Results}

\begin{figure}[h!]
\begin{center}
\includegraphics[width = 7cm]{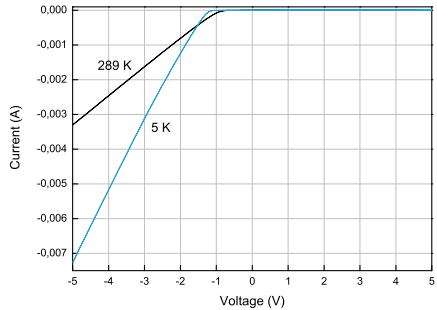}
\caption{I-V characteristics of the junction measured at room temperature (black curve) and 5 K (light blue curve) of the 6 nm thick sample.
\label{fig:G434CycleIRT5K}}
\end{center}
\end{figure}
The $I-V$ curves of the three measured junctions exhibit a diode-like behaviour illustrated in Fig. \ref{fig:G434CycleIRT5K}.
The transition is fully reversible and was observed to be stable over tens of cycles.
At low temperature, the change of resistance with the sign of the voltage is significantly enhanced.

\begin{figure}[h!]
\begin{center}
\includegraphics[width = 7cm]{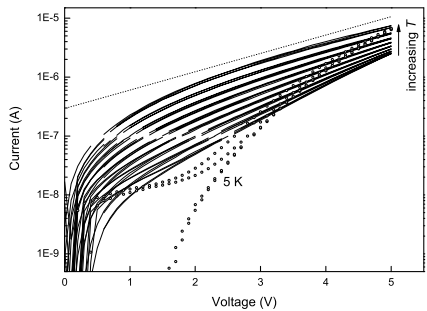}
\caption{Log representation of the current flowing through the 6 nm thick junction in reverse bias at different temperatures from 290 K down to 175 K by steps of 15 K.
The curve measured at 5 K is represented in open circle.
\label{G434IvsVCoolINS}}
\end{center}
\end{figure}
For reverse bias, the current increases with the voltage following an exponential trend (figure \ref{G434IvsVCoolINS}).
As the junctions are cooled down, the $I-V$ curves in log scale are shifted toward higher resistance.
Those properties can be coherently modelled by a tunnel junction in series with an insulator.
Below 175 K, the curves progressively deviate from the high temperature behaviour.
This deviation can be accounted for by a thermoelectric current : 
each copper wire of the voltage probes makes a heterogeneous contact between distinct conductors at distinct temperatures.
As the circuit is closed by the ammeter and the junction cooled down, a thermocouple current adds to the conduction current induced by the applied voltage.
In reverse bias, where the conduction current is vanishing at low temperature, the contribution of the thermoelectric current to the measured current is expected to be significant.
This spurious effect renders the low temperature data little reliable and the discussion will be restricted to the data above 175 K where the temperature gradient is moderate.

\begin{figure}[h!]
\begin{center}
\includegraphics[width = 7cm]{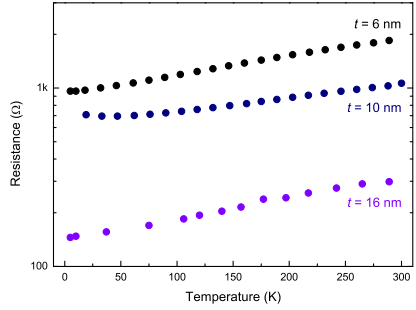}
\caption{Two-wire resistance across the junction as a function of the temperature for three different \LNO{} layer thickness ($t = 6, 10$ and 16 nm) at the applied voltage $\Delta V = -3$ V.}
\end{center}
\end{figure}
The electrical properties in direct bias are represented by the temperature dependence of the resistance measured at the applied voltage $\Delta V = -3$ V.
The residual resistivity ratios range between 1.6 and 2.0 and the estimated resistivity upper bounds (assuming conductive path is about 1 mm long, 3 mm wide and at most as thick as the \LNO{} layer) are 3000 $\mu \Omega \cdot$cm for the 6 and 10 nm samples and 1000 $\mu \Omega \cdot$cm for the 16 nm sample.
Those values are coherent with the literature and one order magnitude above the most conducting single crystals \cite{guo2018} and thick films \cite{king2014}.
Also, the resistance of the junction indirect bias agrees well with that of the \LNO{} film away from the Al over-layer.

\begin{figure}[h!]
\begin{center}
\includegraphics[width = 7cm]{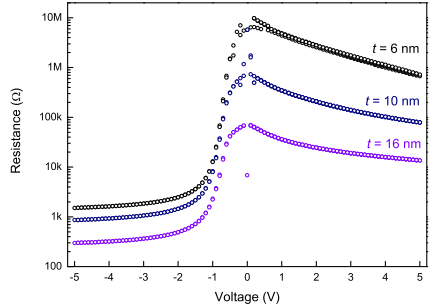}
\caption{Resistance plotted against the bias voltage for three different \LNO{} layer thickness ($t = 6, 10$ and 16 nm) at room temperature.
\label{fig:GThickDependence}
}
\end{center}
\end{figure}
\begin{figure}[h!]
\begin{center}
\includegraphics[width = 7cm]{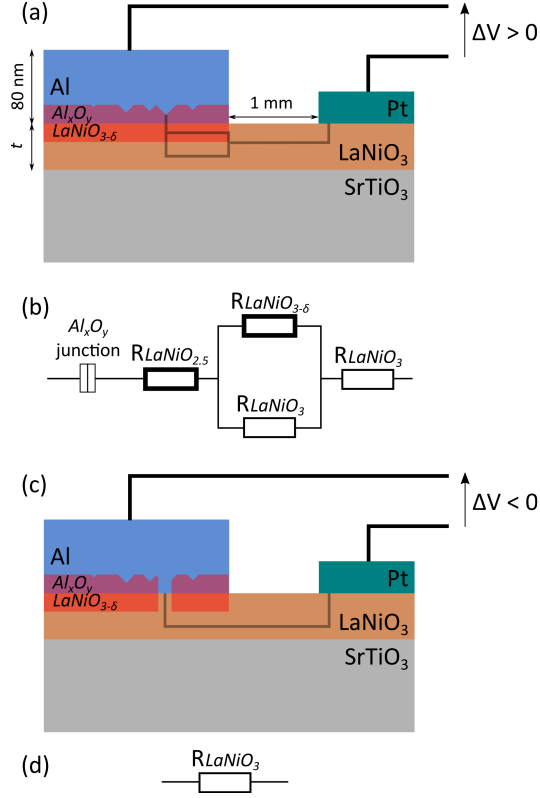}
\caption{Sketches of the conductive path for reverse (a) and direct bias (c) and their respective equivalent circuit (b) and (d).
\label{fig:ConductiveScheme}
}
\end{center}
\end{figure}

The thickness dependence of the electro-resistive effect is reported in Fig. \ref{fig:GThickDependence}.
Electro-resistive effect decreases as the \LNO{} layer increases.
More precisely, the resistance in the reverse bias is much more enhanced than in direct bias as the thickness is diminished.
This trend shows that the oxygen vacancies created by the oxidization of Al are concentrated at the interface:
the thicker the \LNO{} layer is, the thicker is the conducting path through metallic, \ie{} stoichiometric, \LNO{}.
This is coherent with Tian \etal{} who demonstrated that Ni valence is reduced below 3+ only in the first 10 nm of \LNO{} below the interface.
The conducting path for reverse bias can be sketched by a tunnel junction in series with two resistors in parallel representing the semiconducting \LNOo{} and the metallic \LNO{} layers (Fig. \ref{fig:ConductiveScheme}-a and b).

In contrast, the electric behaviour in the direct bias is metallic.
It is as though conducting channels opened through the AlO$_x$/LaNiO$_{2.5}$ bi-layer by a local healing of the Al/\LNO{} interface (Fig. \ref{fig:ConductiveScheme}-c and d).

\section{Discussion}
In this section, the results will be discussed and arguments will be brought to justify the interpretation of the electro-resistive effect in terms of redox reaction controlled by the external electric field.
	
The shape of the $I-V$ curve corresponds to a typical Schottky diode characteristics.
However, in this model, the resistance in the reverse bias is expected to grow, or at least not to exponentially decrease, with the voltage.
Aluminium is well-known to be a strong reducing agent that pumps oxygen out of the oxide it is deposited on \cite{fu2007}.
A direct evidence of the concomitant oxidization of Al and reduction of \LNO{} at a Al/\LNO{} contact have been brought in Ref. \cite{tian2017}.
Thus, the situation can not be described as a regular Schottky contact because of the mass transport which is not considered in this model.

As the Al-\LNO{} contact is formed, electrons are driven from the metal to the oxide by the gradient of chemical potential since $\phi_\mathrm{LNO} > \phi_\mathrm{Al}$ \cite{hong2015, jacobs2016}.
The electrons diffusing in the oxide combine with mobile holes and fill the nearly filled Ni$3d$-O$2p$ band that crosses the Fermi level in \LNO{} \cite{khomskii2014}.
The oxide is turned a semiconductor.
The states the extra electrons set in have been calculated to have an oxygen character \cite{malashevich2015}.
This favours the mobility of oxygen ions.
Thus, the charge carried by the diffusion electrons drives O$^{2-}$ ions outward the oxide where a vacancy is formed and toward the metal where an Al$_x$O$_y$ aluminium oxide layer is formed.
The oxygen transport through the oxide and across the interface is favoured by 
(i) the high mobility of oxygen in \LNO{} (activation barrier of oxygen is 0.65-0.80 eV depending of the tensile strain \cite{mayeshiba2015}),
(ii) the thermodynamic stability of oxygen off-stoichiometric phases
LaNiO$_{2.75}$ and LaNiO$_{2.5}$ \cite{crespin1983, sanchez1996}.
Such formation and spontaneous arrangement of oxygen vacancies have been shown to be an energetically favourable response of the lattice in presence of electrostatic charge \cite{tung2017}.

Finally, the contact between a metal and a metallic oxide makes a highly resistive junction at the equilibrium because of the spontaneous formation of two stable, neutral and insulating layers of Al$_x$O$_y$ and LaNiO$_{3-\delta}$ between the \LNO{} and the Al layers.

\begin{figure}
\begin{center}
\includegraphics[width = 7 cm]{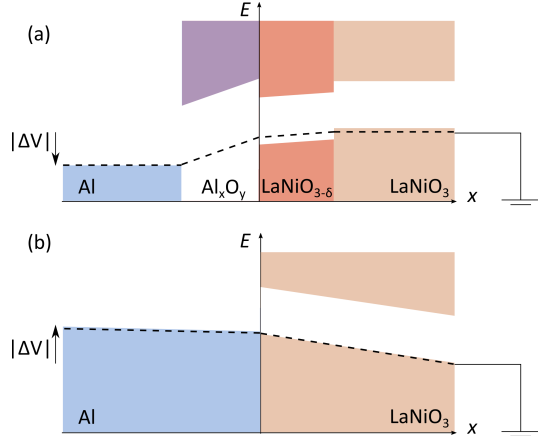}
\caption{
\label{fig:BandDiagrams}
Scheme of the band diagrams across the interface in reverse (a) and direct bias (b).}
\end{center}
\end{figure}
The band diagrams for reverse and direct bias can be sketched as in Fig. \ref{fig:BandDiagrams}.
In reverse bias, mobile charge carriers are strongly suppressed by the opening of a band gap in the intermediate compounds formed at the interface by the redox reaction.
Furthermore, the ionic conduction is impeded by the absence (i) of mobile oxygen ions in \LNOo{} and (ii) of available oxygen sites in Al$_x$O$_y$.
The response of the junction to the applied voltage in reverse bias is the bending the bands in Al$_x$O$_y$ and to some extent in \LNOo{} (figure \ref{fig:BandDiagrams}-a).
In this scheme, the reverse bias can be described as the tunnelling of carriers through the Al$_x$O$_y$/LaNiO$_{3-\delta}$ barrier whose effective height of the barrier decreases as the applied voltage increases.
Its width may be smaller than the average thickness of the Al$_x$O$_y$/LaNiO$_{3-\delta}$ bi-layer due to interface roughness.
From transmission electron microscopy led on analogue systems in Ref. \cite{tian2017}, the Al$_x$O$_y$ thickness can be estimated to be about 10 nm, with enough roughness for weak spots as thin as a few nanometres can exist.
The two compounds yielded by the solid phase redox reaction should have defects so that multiple tunnelling through those defects is likely to take place too.
This would lower the effective width of the barrier and explain why the $I-V$ curves depart from the behaviour of a simple tunnel junction.
In this context, it is difficult to precisely estimate the active area.
The rather surprising reversibility of the electro-resistance process suggests that only little mass is transported, and hence that the area is relatively small.

In direct bias, the applied electric field drives O$^{2-}$ ions from the aluminium oxide toward the nickelate as well as the free holes from \LNO{} toward the \LNOo{} layer.
The holes injected in \LNO{} combine with nickel $d$ electrons and promote Ni host valence to 3+.
This creates available oxygen sites where O$^{2-}$ ions will be driven to satisfy the octahedral coordination of Ni$^{3+}$ sites.
This two-fold transfer of charge and mass turns out to reverse the redox reaction by reducing the aluminium oxide and re-oxidizing \LNO{}.
Above a threshold voltage, the metallic behaviour is recovered all along the conducting path.
The insulator-to-metal transition occurs over a 2 V-wide interval of applied voltages centred around a threshold value of 0.7-0.8 V (resp. 0.8-0.9 V) at room temperature (resp. at 5 K), taking the criterion of a division by 10 of the maximum resistance.

\section{Conclusions}

We reported the electrical behaviour of a Al/\LNO{} contact. 
Both charge and mass transfer take place in this metal-oxide contact and the Schottky effect is strongly enhanced.
Due to the occurrence of a redox reaction, insulating intermediate compounds are formed at the interface between those two metallic native compounds and the junction is highly resistive at the equilibrium.
The junction exhibits either tunnelling-like conduction or a metallic behaviour depending on the sign of the applied voltage.
The fabrication of such a junction is straightforward with basic growth tools, inexpensive, efficient at room temperature.
On top of that, the mass transfer is found to be confined to a few nanometres around the interface.
All these features makes this system promising for applications in nano-electronics.

\section{Acknowledgement}
This work was supported by the Ile-de-France region for magneto-transport measurements ("NOVATECS" C'Nano IdF project n$^\circ$ IF-08-1453/R)

\bibliography{RedoxSchottky}

\end{document}